\documentclass[10pt,sort&compress]{article}
\usepackage{graphicx}
\usepackage{bm,multirow}
\usepackage{amsmath,amssymb,color,mathrsfs}
\usepackage{footnote}
\usepackage{citesort}
\usepackage{cuted}

\def\lambdabar{\lambda\kern-1ex\raise0.55ex\hbox{--}}

\newcommand{\AJM}{Am. J. Math. }
\newcommand{\AJP}{Am. J. Phys. }
\newcommand{\AM}{Annals Math. }

\newcommand{\APB}{Ann. Phys. (Berlin) }
\newcommand{\APNY}{Ann. Phys. (N.Y.) }

\newcommand{\CMP}{Commun. Math. Phys. }

\newcommand{\CPL}{Chem. Phys. Lett. }
\newcommand{\CRA}{C. R. Acad. Sci. Ser. A }
\newcommand{\EJP}{Eur. J. Phys. }
\newcommand{\EPJD}{Eur. Phys. J. D }
\newcommand{\IJMPB}{Int. J. Mod. Phys. B }

\newcommand{\IJQC}{Int. J. Quantum Chem. }

\newcommand{\JCP}{J. Chem. Phys. }

\newcommand{\JMP}{J. Math. Phys. }
\newcommand{\jpa}{J. Phys. A }

\newcommand{\NJP}{New J. Phys. }
\newcommand{\NL}{Nature (London) }

\newcommand{\PLA}{Phys. Lett. A }

\newcommand{\PRA}{Phys. Rev. A }
\newcommand{\PRB}{Phys. Rev. B }
\newcommand{\PRD}{Phys. Rev. D }
\newcommand{\PRE}{Phys. Rev. E }
\newcommand{\PRL}{Phys. Rev. Lett. }

\definecolor{officegreen}{rgb}{0,0.5,0}
\definecolor{pakistangreen}{rgb}{0,0.4,0}
\definecolor{palatinatepurple}{rgb}{0.41,0.16,0.38}
\definecolor{sangria}{rgb}{0.57,0,0.04}

\begin{document}
\title{Theory of the Robin quantum wall in a linear potential. I. Energy spectrum, polarization and quantum-information measures}
\author{O. Olendski\footnote{Department of Physics, King Saud University, P.O. Box 2455, Riyadh 11451 Saudi Arabia; E-mail: oolendski@ksu.edu.sa}}

\maketitle

\begin{abstract}
Information-theoretical concepts are employed for the analysis of the interplay between a transverse electric field $\mathscr{E}$ applied to a one-dimensional surface and Robin boundary condition (BC), which with the help of the extrapolation length $\Lambda$ zeroes at the interface a linear combination of the quantum mechanical wave function and its spatial derivative, and its influence on the properties of the structure. For doing this, exact analytical solutions of the corresponding Schr\"{o}dinger equation are derived and used for calculating energies, dipole moments, position $S_x$ and momentum $S_k$ quantum information entropies and their Fisher information $I_x$ and $I_k$ and Onicescu information energies $O_x$ and $O_k$ counterparts. It is shown that the weak (strong) electric field changes the Robin wall into the Dirichlet, $\Lambda=0$ (Neumann, $\Lambda=\infty$), surface. This transformation of the energy spectrum and associated waveforms in the growing field defines an evolution of the quantum-information measures; for example, it is proved that for the Dirichlet and Neumann BCs the position (momentum) quantum information entropy varies as a positive (negative) natural logarithm of the electric intensity what results in their field-independent sum $S_x+S_k$. Analogously,  at $\Lambda=0$ and $\Lambda=\infty$ the position and momentum Fisher informations (Onicescu energies) depend on the applied voltage as $\mathscr{E}^{2/3}$ ($\mathscr{E}^{1/3}$) and its inverse, respectively, leading to the field-independent product $I_xI_k$ ($O_xO_k$). Peculiarities of their transformations at the finite nonzero $\Lambda$ are discussed and similarities and differences between the three quantum-information measures in the electric field are highlighted with the special attention being paid to the configuration with the negative extrapolation length.
\end{abstract}

\section{Introduction}\label{Introduction}
Properties of the physical objects are strongly affected by the confining surfaces. To describe their influence, theoretical physics supplements the wave equation, which we will write in the stationary Schr\"{o}dinger-like form
\begin{equation}\label{Schrodinger1}
-\frac{\hbar^2}{2m}{\bm\nabla}^2\Psi({\bf r})+V({\bf r})\Psi({\bf r})=E\Psi({\bf r}),
\end{equation}
by the boundary condition (BC) for the scalar field $\Psi({\bf r})$ at the interface $\cal S$:
\begin{equation}\label{Robin1}
\left.{\bf n}{\bm\nabla}\Psi\right|_{\cal S}=\left.\frac{1}{\Lambda}\Psi\right|_{\cal S}.
\end{equation}
Here, $m$ is a mass of the particle (for definiteness, we will talk about the electron), $E$ is its energy, $V({\bf r})$ is an external potential, $\bf n$ is an inward unit vector normal to the confining wall. Linear relation between the function $\Psi$ and its spatial derivative is governed by the Robin length $\Lambda$ \cite{Gustafson1} whose real value guarantees that no current with the density
\begin{equation}\label{CurrentDensity1}
{\bf j}=-\frac{e\hbar}{m}{\rm Im}(\Psi^*{\bm\nabla}\Psi)
\end{equation}
($e$ is an absolute value of the electronic charge) flows through the surface:
\begin{equation}\label{CurrentDensity2}
\left.{\bf nj}\right|_{\cal S}\equiv0\quad{\rm at}\quad{\rm Im}(\Lambda)=0.
\end{equation}
Note that the same Eq.~\eqref{Schrodinger1} (with the appropriate change of its parameters) and the BC from Eq.~\eqref{Robin1} describe the processes in acoustics \cite{Felix1}, electrodynamics \cite{Katsenelenbaum1}, plasma \cite{Silva1}, scalar field theory \cite{Solodukhin1,Saharian1,Romeo1,Elizalde1}, superconductivity \cite{Fink1,Montevecchi1,Kozhevnikov1,Giorgi1} where the coefficient $\Lambda$ is called the de Gennes distance \cite{deGennes1}, and others \cite{Sapoval1,Essert1}. Nonexhaustive list of the relevant research efforts can be found in Refs.~\cite{Olendski1,Olendski2,Olendski2_1,Grebenkov1}.

Varying the positive Robin distance from its zero value to the infinity allows a continuous change of the BC from the Dirichlet, $\Psi|_{\cal S}=0$, to the Neumann one, ${\bf n}{\bm\nabla}\Psi|_{\cal S}=0$, respectively. The most interesting is the situation with the negative extrapolation length $\Lambda$. Consider, for example, one-dimensional (1D) motion of the electron in the potential-free region $x\le0$, which is confined from the right by the Robin wall \cite{Seba1,Pazma1,Fulop1,Belchev1,Georgiou1}. For nonnegative extrapolation parameter, the scattering states with only the positive energies exist while for $\Lambda<0$ there is additionally a bound level with the energy
\begin{equation}\label{SingleRobinWallEnergy1}
E=-\frac{\hbar^2}{2m\Lambda^2},
\end{equation}
whose normalized to unity,
\begin{equation}\label{Normalization1}
\int_{-\infty}^0\Psi^2(x)dx=1,
\end{equation}
wave function $\Psi(x)$ exponentially vanishes at the negative infinity:
\begin{equation}\label{FunctionPsi1}
\Psi(x)=\left(\frac{2}{|\Lambda|}\right)^{1/2}\!\!\exp\!\left(\frac{x}{|\Lambda|}\right),\quad x\le0.
\end{equation}
Its momentum counterpart
\begin{equation}\label{FunctionPhi1}
\Phi(k)=\frac{1}{\left(2\pi\right)^{1/2}}\int_{-\infty}^0e^{-ikx}\Psi(x)dx
\end{equation}
obeys the normalization
\begin{equation}\label{Normalization2}
\int_{-\infty}^\infty|\Phi(k)|^2dk=1
\end{equation}
and is given as
\begin{equation}\label{FunctionPhi2}
\Phi(k)=\left(\frac{|\Lambda|}{\pi}\right)^{1/2}\frac{1}{1-i|\Lambda|k}.
\end{equation}
Eqs.~\eqref{SingleRobinWallEnergy1} and \eqref{FunctionPsi1} show that the wall with the negative Robin length acts as an attractive surface: the smaller the absolute value $|\Lambda|$ is, the stronger the attraction and localization of the particle at the interface are. The limit $\Lambda\rightarrow-0$ is the point of non-analyticity of the system. Experimentally, the structures with the negative extrapolation distances were fabricated by using superconductors \cite{Fink1,Kozhevnikov1}. It was also argued that the 1D wall with $\Lambda<0$ can be realized as a limit of finite regularized potentials \cite{Pazma1,Fulop1,AlHashimi1} that can be readily manufactured using, e.g., thin layers of different types of semiconductors.

Position $\Psi(x)$ and momentum $\Phi(k)$ wave functions through the corresponding densities $\rho(x)$ and $\gamma(k)$
\begin{subequations}\label{densityXK_1}
\begin{eqnarray}\label{densityX_1}
\rho(x)=|\Psi(x)|^2\\
\label{densityK_1}
\gamma(k)=|\Phi(k)|^2
\end{eqnarray}
\end{subequations}
define associated moments $\left<x^n\right>$ and $\left<k^n\right>$, $n=1,2,\ldots$, respectively:
\begin{subequations}\label{XKaveraging1}
\begin{eqnarray}\label{Xaveraging1}
\left<x^n\right>=\int_{-\infty}^0 x^n\rho(x)dx\\
\label{Kaveraging1}
\left<k^n\right>=\int_{-\infty}^\infty k^n\gamma(k)dk.
\end{eqnarray}
\end{subequations}
Two lowest momenta of $x$ and $k$ enter the Heisenberg uncertainty relation
\begin{equation}\label{Heisenberg1}
\Delta x\Delta k\ge\frac{1}{2}
\end{equation}
with $\Delta x$ and $\Delta k$ being position and momentum standard deviations:
\begin{subequations}\label{DeltaXK1}
\begin{eqnarray}\label{DeltaX1}
\Delta x=\sqrt{\left<x^2\right>-\left<x\right>^2}\\
\label{DeltaK1}
\Delta k=\sqrt{\left<k^2\right>-\left<k\right>^2}.
\end{eqnarray}
\end{subequations}
It is immediately seen that the Heisenberg inequality \eqref{Heisenberg1} for the state defined by Eqs.~\eqref{SingleRobinWallEnergy1}, \eqref{FunctionPsi1} and \eqref{FunctionPhi2} is meaningless since the second-order moment $\left<k^2\right>$ diverges. A situation is getting even worse if one calculates the momenta over the position space where $k$ is an operator, $\hat{k}=-i\partial_x$, acting upon the functions $\Psi(x)$ what in our case results in $\Delta k=0$ obviously violating Eq.~\eqref{Heisenberg1}. Explanation of this seeming paradox lies in the fact that the Heisenberg uncertainty  assumes that the wave function $\Psi$ vanishes at infinity \cite{AlHashimi1} what is not the case for the Robin wall. To eliminate this discrepancy, one introduces the BC dependent terms into the uncertainty relation \cite{AlHashimi1}. Another option is to use some different information measures for the description of the system. Probably, the most popular among contemporary physicists for doing this is a quantum information entropy. Introduced initially by C. E. Shannon for the mathematical analysis of communication \cite{Shannon1}, it is attracting more and more attention from the researchers of the nano world \cite{Gadre1,Angulo1,Yanez1,Aptekarev1,Dehesa1,Majernik1,Majernik2,Majernik3,Majernik4,Massen1,Sun1,Laguna1,Olendski3,Mukherjee1,Mukherjee2}. Quantum information entropy can be defined both in the position $S_x$ as well as momentum $S_k$ space, which for our 1D system under consideration read:
\begin{subequations}\label{Entropy1}
\begin{align}\label{EntropyX_1}
S_x=-\int_{-\infty}^0\rho(x)\ln\rho(x)dx\\
\label{EntropyK_1}
S_k=-\int_{-\infty}^{\infty}\gamma(k)\ln\gamma(k)dk.
\intertext{
These functionals are objective measures of the uncertainty or missing information of the corresponding distributions. It was first proved by Beckner \cite{Beckner1} and Bia{\l}ynicki-Birula and Mycielski \cite{Bialynicki1} (see also earlier conjectures \cite{Everett1,Hirschman1}) that for the general 1D configuration their sum}
\label{EntropyTotal1}
S_t=S_x+S_k
\end{align}
\end{subequations}
satisfies the uncertainty
\begin{equation}\label{EntropicInequality1}
S_t\ge1+\ln\pi.
\end{equation}
Elementary calculation yields for the Robin wall \cite{Olendski3}:
\begin{subequations}\label{SingleRobinWallEntropy}
\begin{align}\label{SingleRobinWallEntropyX}
S_x&=1-\ln2+\ln|\Lambda|\\
\label{SingleRobinWallEntropyK}
S_k&=2\ln2+\ln\pi-\ln|\Lambda|,
\intertext{what results in the $\Lambda$ independent finite sum:}
\label{SingleRobinWallEntropySnoFields}
S_t&=1+\ln\pi+\ln2,
\end{align}
\end{subequations}
satisfying, of course, Eq.~\eqref{EntropicInequality1}. This simple comparative example is another confirmation of the fact \cite{Deutsch1,Partovi1} that entropic uncertainty relation, Eq.~\eqref{EntropicInequality1}, is stronger than its Heisenberg counterpart, Eq.~\eqref{Heisenberg1}, as it presents more general base for defining 'uncertainty' \cite{Bialynicki2,Wehner1}.

Another quantitative measure of entanglement is provided by the Fisher information. Proposed by statistician and geneticist about the same time when the quantum mechanics was born \cite{Fisher1}, it  finds more and more applications not only in physics but in other miscellaneous branches of science \cite{Frieden1}. Similar to entropy, it can be defined in position and momentum spaces. Their general $n$-dimensional definitions
\begin{subequations}\label{GeneralFisher1}
\begin{eqnarray}\label{GeneralFisherX1}
I_{\bf r}&=&\int_{\mathbb{R}^n}\rho({\bf r})\left|{\bm\nabla}\ln\rho({\bf r})\right|^2d^n{\bf r}=\int_{\mathbb{R}^n}\frac{\left|{\bm\nabla}\rho({\bf r})\right|^2}{\rho({\bf r})}d^n{\bf r}\\
\label{GeneralFisherK1}
I_{\bf k}&=&\int_{\mathbb{R}^n}\gamma({\bf k})\left|{\bm\nabla}\ln\gamma({\bf k})\right|^2d^n{\bf k}=\int_{\mathbb{R}^n}\frac{\left|{\bm\nabla}\gamma({\bf k})\right|^2}{\gamma({\bf k})}\,d^n{\bf k}
\end{eqnarray}
\end{subequations}
for our 1D system simplify to
\begin{subequations}\label{Fisher1D_1}
\begin{eqnarray}\label{Fisher1D_X1}
I_x=\int_{-\infty}^0\rho(x)\left[\frac{d}{dx}\ln\rho(x)\right]^2dx=\int_{-\infty}^0\frac{\rho'(x)^2}{\rho(x)}\,dx\\
\label{Fisher1D_K1}
I_k=\int_{-\infty}^\infty\gamma(k)\left[\frac{d}{dk}\ln\gamma(k)\right]^2dk=\int_{-\infty}^\infty\frac{\gamma'(k)^2}{\gamma(k)}dk.
\end{eqnarray}
\end{subequations}
Presence of the gradient, which measures the speed of change of the corresponding density, makes the Fisher information a local measure of uncertainty while the quantum entropy containing the logarithm is a global mapping of the charge distribution. Stronger position or momentum localization of the electron means higher value of the corresponding Fisher information. Contrary to the quantum entropies, which always obey the lower bound from Eq.~\eqref{EntropicInequality1}, similar universal relation between position and momentum components of the Fisher information is not known though for several particular systems some inequalities have been derived \cite{Stam1,Dembo1,Romera1,Dehesa2,Dehesa3}. In quantum mechanics, one of its main applications is due to the fact that it enters into the expression for the kinetic energy of the many-particle system \cite{Sears1} and in this way establishes the link between, for example, density functional methods and information theory. Due to its importance, a research on the Fisher information and its relation to the quantum entropy is a very vigorous one and it discovers a lot of the new results \cite{Mukherjee1,Romera1,Romera2,Nagy1,Dehesa4,Nagy2,Sen1,LopezRosa1,Toranzo1,Nagy3,Macedo1,Falaye1}. For our subsequent analysis,  we provide here easily derivable expressions for the position and momentum Fisher informations of the bound state of the negative Robin wall:
\begin{subequations}\label{FisherRobin1}
\begin{align}\label{FisherRobinX1}
I_x=\frac{4}{|\Lambda|^2}\\
I_k=\frac{|\Lambda|^2}{2}.
\intertext{These equations show that the product of the two informations is a BC-independent constant:}
\label{FisherRobinProduct1}
I_xI_k=2.
\end{align}
\end{subequations}

Among other quantum-theoretical measures, let us mention also information energy introduced in 1966 by O. Onicescu \cite{Onicescu1}. For our system, its position $O_x$ and momentum $O_k$ components are:
\begin{subequations}\label{Onicescu1}
\begin{eqnarray}\label{Onicescu1_X}
O_x=\int_{-\infty}^0\rho^2(x)dx\\
\label{Onicescu1_K}
O_k=\int_{-\infty}^\infty\gamma^2(k)dk.
\end{eqnarray}
\end{subequations}
In other words, Onicescu energies are the mean values of the corresponding probability densities or quadratic deviations from the probability equilibria. To underline the difference between the quantum entropy and information energy, let us consider a simple example that justifies the study of the latter; namely, it is easy to show \cite{Lepadatu1,Chatzisavvas1} that for the discrete field with $N$ events the information energy (entropy) reaches minimum of $1/N$ (maximum of $\ln N$) when the likelihoods of all occurences are equal while the unit maximum (zero minimum) takes place with the probability of one event being certain with all others turning to zeros. Since the former case corresponds to a complete disorder, by analogy with thermodynamics the quantity $O$ is coined as 'energy' though actually it is measured in units of the inverse volume of the field upon which it is calculated. Similar to the quantum entropy and Fisher information, it finds applications not only in physics but, for example, in social sciences \cite{Avram1}. A comparison of the three information measures has been performed for a number of systems \cite{Mukherjee1,Lepadatu1,Chatzisavvas1,Avram1,Agop1}. For the only bound state of the negative Robin wall the information energies are:
\begin{subequations}\label{Onicescu2}
\begin{align}\label{Onicescu2_X}
O_x=\frac{1}{|\Lambda|}\\
\label{Onicescu2_K}
O_k=\frac{|\Lambda|}{2\pi},\\
\intertext{what, similar to Fisher information, makes their dimensionless product a $\Lambda$ independent quantity:}
\label{Onicescu2_XK}
O_xO_k=\frac{1}{2\pi}.
\end{align}
\end{subequations}

For the example from the previous paragraph it is seen that the product of the Onicescu energy and the exponent of the quantum entropy 
\begin{equation}\label{CGLdefinition1}
CGL=e^SO
\end{equation}
stays equal to unity in the extreme limits of completely ordered and totally disordered structure since these multipliers change in the opposite directions. Statistical measure of complexity from Eq.~\eqref{CGLdefinition1} was introduced by Catal\'{a}n, Garay and L\'{o}pez-Ruiz \cite{Catalan1} to avoid the shortcomings \cite{Feldman1} of the multiplication of the entropy and information energy $SO$ \cite{LopezRuiz1}. Physically, the product from Eq.~\eqref{CGLdefinition1} scales the speeds of change of the entropy and information energy to the same order of magnitude. As a result, for the bound state of the attractive Robin wall its position and momentum components do not depend on the extrapolation length:
\begin{subequations}\label{CGL1}
\begin{eqnarray}\label{CGL1_X}
CGL_x&=&\frac{e}{2}\\
\label{CGL1_K}
CGL_k&=&2\\
\label{CGL1_XK}
CGL&=&CGL_x\cdot CGL_k=e,
\end{eqnarray}
\end{subequations}
as it follows from Eqs.~\eqref{SingleRobinWallEntropy} and \eqref{Onicescu2}. This measure was used, for example, for studying neutron star \cite{Chatzisavvas2,deAvellar1} and white dwarf \cite{Sanudo3} structure. It was also calculated for the isotropic 3D harmonic oscillator \cite{Sanudo1} and the hydrogen atom \cite{Sanudo2}.

Returning to Eq.~\eqref{Schrodinger1}, let us point out that the potential $V({\bf r})$ in it describes the influence of the external sources on the properties of the system; for example, uniform electric field $\mathscr{E}$ directed in the positive $x$ direction is created  by the associated potential $V(x)=-e\mathscr{E}x$ while for the charged particle in the gravitational field it becomes $V(x)=-mgx$, with $g$ being the acceleration in the Earth's gravitational field. The interest in the latter configuration together with the horizontal Dirichlet mirror, $\Psi(x=0)=0$, coined as a 'quantum bouncer' \cite{Gibbs1}, has been  renewed after the impressive experimental measurement of the quantum states of neutrons in the Earth's gravitational field \cite{Nesvizhevsky1} what resulted, in particular, in the experimental observation of a photon bouncing ball \cite{DellaValle1}. It is natural to wonder: how the wall with the nonzero extrapolation length, $\Lambda\ne0$, will affect the properties of the particle in the linear potential? Previous calculations of the hydrogen in parallel static magnetic and electric fields \cite{GonzalezFerez1} and Rydberg potassium atom in the electric field \cite{He1} showed that the knowledge of the changes of position Shannon information entropy with the field is a good tool for prediction of their characteristics.

In the present research, a comprehensive quantum analysis of the electron behavior in the presence of the Robin wall with the external electric field $\mathscr{E}$ that pushes it to the surface is carried out. Analytical solutions $\Psi_n(x)$ of the Schr\"{o}dinger equation lead to the transcendental expression for calculating energies $E_n$ as a function of the applied voltage. It is proved that for the weak fields the spectrum reduces to the one corresponding to the Dirichlet wall while for the strong electric intensities it transforms into the Neumann one. The knowledge of the energies and waveforms  allows to calculate and analyze, in addition to the quantities discussed above, a polarization, or dipole moment, $P$ expression for which reads \cite{Nguyen1,Olendski1}:
\begin{equation}\label{Polarization1}
P(\mathscr{E})=\left\langle ex\right\rangle_\mathscr{E}-\left\langle ex\right\rangle_{\mathscr{E}=0},
\end{equation}
where the angle brackets denote a quantum mechanical expectation value:
\begin{equation}\label{AngleBrackets1}
\left\langle x\right\rangle=\int_{-\infty}^0x\Psi^2(x)dx.
\end{equation}
Polarization is a quantitative measure of the charge redistribution in the electric field. In a broader context, it presents diagonal elements of the dipole moment matrix $\bf P$ with elements
\begin{equation}\label{PolarizMatrix1}
P_{nm}=e\int_{-\infty}^0x\Psi_n(x)\Psi_m(x)dx.
\end{equation}
Knowledge of the nondiagonal elements is crucial for, e.g., description of the optical transitions between states with the quantum numbers $n$ and $m$ \cite{Ahn1}.

The paper is organized as follows. In Sec.~\ref{Sec_Energy} a discussion of the energy spectrum and polarization matrix
is carried out. Sec.~\ref{Sec_QuantumInfo} is devoted to the study of the quantum-information measures and their comparative analysis with separate subsections on the quantum information entropy, Fisher information and Onicescu energy (together with the complexity $CGL$). Similarities and differences between these quantities are underlined. The presentation is wrapped up with some conclusions in Sec.~\ref{Sec_Conclusions}. Similar to the quantum well with miscellaneous combinations of the Dirichlet and Neumann BCs at the opposite walls  in the electric field \cite{Olendski3,Olendski4}, a companion paper \cite{Olendski5} calculates statistical properties of the same structure.

\section{Energy Spectrum and Dipole Moment}\label{Sec_Energy}
A starting point of our analysis is the 1D Schr\"{o}dinger equation
\begin{equation}\label{Schrodinger2}
\hat{H}\Psi(x)=E\Psi(x)
\end{equation}
with the Hamiltonian $\hat{H}$
\begin{equation}\label{Hamiltonian1}
\hat{H}=-\frac{\hbar^2}{2m}\frac{d^2}{dx^2}-e\mathscr{E}x
\end{equation}
for the wave function $\Psi(x)$ of the charged particle moving on the half-line $-\infty<x\le0$ in the uniform electric field $\mathscr{E}$ with the BC at the edge $x=0$ being of the form:
\begin{equation}\label{Robin2}
-\Lambda\Psi'(0)=\Psi(0)
\end{equation}
with the prime denoting a derivative of the function with respect to its argument. For the gravitational field $g$, in all equations the expression $-e\mathscr{E}$ has to be replaced by $-mg$. Below, we will use the superscript D (N) for denoting Dirichlet, $\Lambda=0$ (Neumann, $\Lambda=\infty$), type of the BC at the interface while the character R followed, if necessary, by the plus (minus) sign will refer to the Robin surface with the positive (negative) extrapolation length. It is convenient from the very beginning to switch to the dimensionless units. For the finite nonzero Robin length the appropriate scaling measures all distances in units of $|\Lambda|$, energies -- in units of $\hbar^2/(2m|\Lambda|^2)$, polarization -- in units of $e|\Lambda|$, and electric fields -- in units of $\hbar^2/(2em|\Lambda|^3)$. As a result, Eqs.~\eqref{Hamiltonian1} and \eqref{Robin2} transform to the universal form:
\begin{eqnarray}\label{Hamiltonian2}
\hat{H}=-\frac{d^2}{dx^2}-\mathscr{E}x\\
\label{Robin3}\pm\Psi'(0)=\Psi(0),
\end{eqnarray}
where the upper (lower) sign refers to the negative (positive) Robin distance. Of course, such scaling fails for the Dirichlet and Neumann BCs. In this case, one can use as a unit of length any appropriate distance, say, the Earth or Sun radius, but for the quantum mechanical applications the most reasonable is the adoption of the reduced Compton wavelength $\lambdabar=\hbar/(mc)$, what naturally leads to the units of energy $mc^2$, and electric field - $m^2c^3/(e\hbar)$, with $c$ being the speed of light. As a result, the Hamiltonian again takes the form of Eq.~\eqref{Hamiltonian2}, and the BC turns to $\Psi(0)=0$ for the Dirichlet wall and $\Psi'(0)=0$ for the Neumann one.

\begin{figure}
\centering
\includegraphics*[width=\columnwidth]{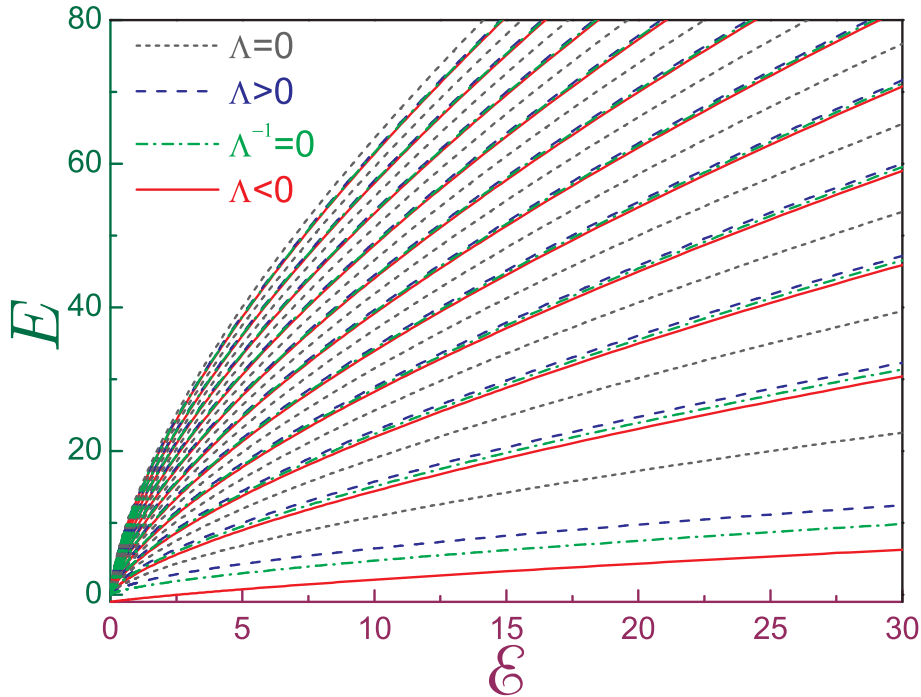}
\caption{\label{EnergySpectrum}
Energy spectrum $E_n$ for the negative (solid curve) and positive (dashed line) Robin length, Dirichlet (dotted curve) and Neumann (dash-dotted line) wall as a function of the electric field $\mathscr{E}$.}
\end{figure}

The structure is highly asymmetric with respect to the sign of the field: for the positive electric intensities the spectrum is completely discrete with the countably infinite number of the bound states while for the negative $\mathscr{E}$ it is continuous. The latter case is considered elsewhere \cite{Olendski6}. Turning to the discussion of the geometry with $\mathscr{E}>0$, one notices that the normalized to unity, Eq.~\eqref{Normalization1}, wave functions $\Psi_n(x)$, $n=0,1,2,\ldots$, of the bound levels
\begin{subequations}\label{WaveFunction1}
\begin{eqnarray}\label{WaveFunction1_Robin}
\Psi_n^R(x)&=&\left(\frac{\mathscr{E}}{E_n+1}\right)^{1/2}\frac{{\rm Ai}\!\left(-\mathscr{E}^{1/3}x-E_n/\mathscr{E}^{2/3}\right)}{{\rm Ai}\!\left(-E_n/\mathscr{E}^{2/3}\right)}\\
\label{WaveFunction1_Dirichlet}
\Psi_n^D(x)&=&\mathscr{E}^{1/6}\frac{{\rm Ai}\!\left(-\mathscr{E}^{1/3}x+a_{n+1}\right)}{{\rm Ai}'\!\left(a_{n+1}\right)}\\
\label{WaveFunction1_Neumann}
\Psi_n^N(x)&=&\mathscr{E}^{1/6}\frac{{\rm Ai}\!\left(-\mathscr{E}^{1/3}x+a_{n+1}'\right)}{\left|a_{n+1}'\right|^{1/2}{\rm Ai}\!\left(a_{n+1}'\right)},
\end{eqnarray}
\end{subequations}
exponentially fade away from the wall: $\Psi(x)\xrightarrow[x\rightarrow-\infty]{}0$. In these equations, ${\rm Ai}(x)$ is Airy function  and negative coefficient $a_n$, $n=1,2,\ldots$, is its $n$th root while $a_n'$ satisfies ${\rm Ai}'(x)=0$ \cite{Abramowitz1,Vallee1}. Energy spectrum $E_n$ is found from the following transcendental equation:
\begin{equation}\label{EigenValueRobin1}
\mathscr{E}^{1/3}{\rm Ai}'\!\!\left(-\frac{E}{\mathscr{E}^{2/3}}\right)\pm{\rm Ai}\!\!\left(-\frac{E}{\mathscr{E}^{2/3}}\right)=0,
\end{equation}
while for the Dirichlet \cite{Gibbs1,Katriel1} and Neumann wall it is given as
\begin{equation}\label{EigenValueDirNeumann1}
E_n^{\left\{_N^D\right\}}=-\mathscr{E}^{2/3}\left\{\begin{array}{cc}a_{n+1}\\a_{n+1}'\end{array}\right\},\quad n=0,1,2,\ldots.
\end{equation}
As it follows from Eq.~\eqref{EigenValueRobin1}, for the strong fields, the energy spectrum reduces basically to the Neumann one:
\begin{equation}\label{AsymptoticNeumann1}
E_n^{R\mp}=-a_{n+1}'\mathscr{E}^{2/3}\left(1\mp\frac{1}{{a_{n+1}'}^{\!\!\!\!\!\!\!\!2}\quad\mathscr{E}^{1/3}}\right),\quad\mathscr{E}\gg1,
\end{equation}
where the second item in the right-hand side is a tiny admixture due to the finiteness of the extrapolation length. For the very weak potentials it is described mainly by the Dirichlet dependence:
\begin{subequations}\label{AsymptoticDirichlet1}
\begin{align}\label{AsymptoticDirichlet1_NP}
E_n^{R\mp}&=-\left\{\!\!\begin{array}{c}
a_n\\a_{n+1}
\end{array}\!\!\right\}\mathscr{E}^{2/3}\pm\mathscr{E},\,\left\{\!\!\begin{array}{c}
n=1,2,3,\ldots\\n=0,1,2,\ldots
\end{array}\!\!\right\},\quad\mathscr{E}\ll1\\
\intertext{with the second term being the higher-order (linear) coefficient of the Taylor expansion with respect to a small parameter $\Lambda$. Additionally, for the negative extrapolation lengths,}\label{AsymptoticDirichlet1_Negative0}
E_0^{R-}&=-1+\frac{1}{2}\,\mathscr{E}-\frac{1}{8}\,\mathscr{E}^2,\quad\mathscr{E}\ll1.
\end{align}
\end{subequations}
Lowest level passes zero energy at the field $\mathscr{E}_{E_0=0}$, which follows from Eq.~\eqref{EigenValueRobin1} and properties of the Airy functions \cite{Abramowitz1,Vallee1}
\begin{equation}\label{BetaZeroEnergy1}
\mathscr{E}_{E_0=0}=\frac{1}{3}\frac{\Gamma^3(1/3)}{\Gamma^3(2/3)}=2.58106\ldots,
\end{equation}
with $\Gamma(x)$ being $\Gamma$-function \cite{Abramowitz1}.

\begin{figure}
\centering
\includegraphics*[width=\columnwidth]{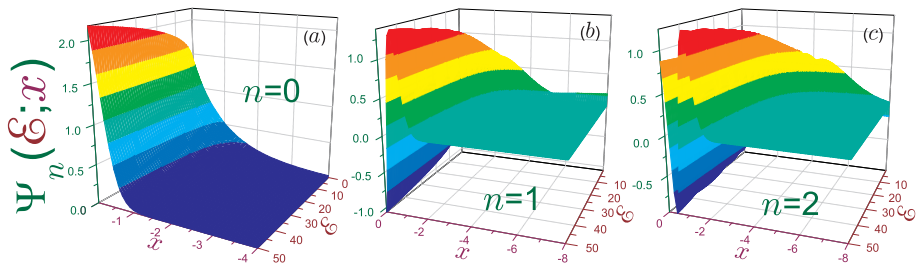}
\caption{\label{Functions}
Function $\Psi(x)$ of the (a) lowest level, (b) first and (c) second field-induced bound states of the negative Robin wall in terms of the distance $x$ and electric intensity $\mathscr{E}$. Note different $x$ scale in panel (a) as compared to other two plots. In addition, each vertical axis has its own range.}
\end{figure}

Several low-lying energies are shown in Fig.~\ref{EnergySpectrum} for the negative (solid line) and positive (dashed curve) extrapolation lengths together with the Dirichlet (dotted line) and Neumann (dash-dotted curve) BCs. It is seen that for the positive energies and small fields the density of states increases  with the decreasing electric intensity; for example, for either positive or negative Robin distances the $n$-dependent difference between the two adjacent levels $\Delta E_n$ is proportional to $\mathscr{E}^{2/3}$:
\begin{equation}\label{DeltaE2}
\Delta E_n=\mathscr{E}^{2/3}(a_n-a_{n+1}),\quad\mathscr{E}\ll1.
\end{equation}
From the asymptotic expansion of the coefficients $a_n$ \cite{Abramowitz1,Vallee1} it follows that the positive difference $\Delta a_n=a_n-a_{n+1}$ is getting smaller for the larger $n$ 
\begin{equation}\label{DeltaAn}
\Delta a_n=\left(\frac{2}{3}\frac{\pi^2}{n}\right)^{1/3},\quad n\gg1,
\end{equation}
what means that at the fixed weak field the number of states per unit energy grows with $E$. In addition to the positive spectrum, for the negative de Gennes length there exists the lower lying negative-energy state separated from the quasi-continuum by the almost unit-energy gap. As will be shown in Ref.~\cite{Olendski5}, such structure of the spectrum leads to spectacular features of the thermodynamic properties. For the strong fields, the energies of the Robin interface approach those of the Neumann wall  with this asymptote being reached faster for the higher lying states. For each fixed $n$, the energy of the negative Robin level is the lowest one followed by the Neumann, positive Robin and Dirichlet wall:
\begin{equation}\label{Sequence1}
E_n^{R-}<E_n^N<E_n^{R+}<E_n^D.
\end{equation}
\begin{figure}
\centering
\includegraphics*[width=\columnwidth]{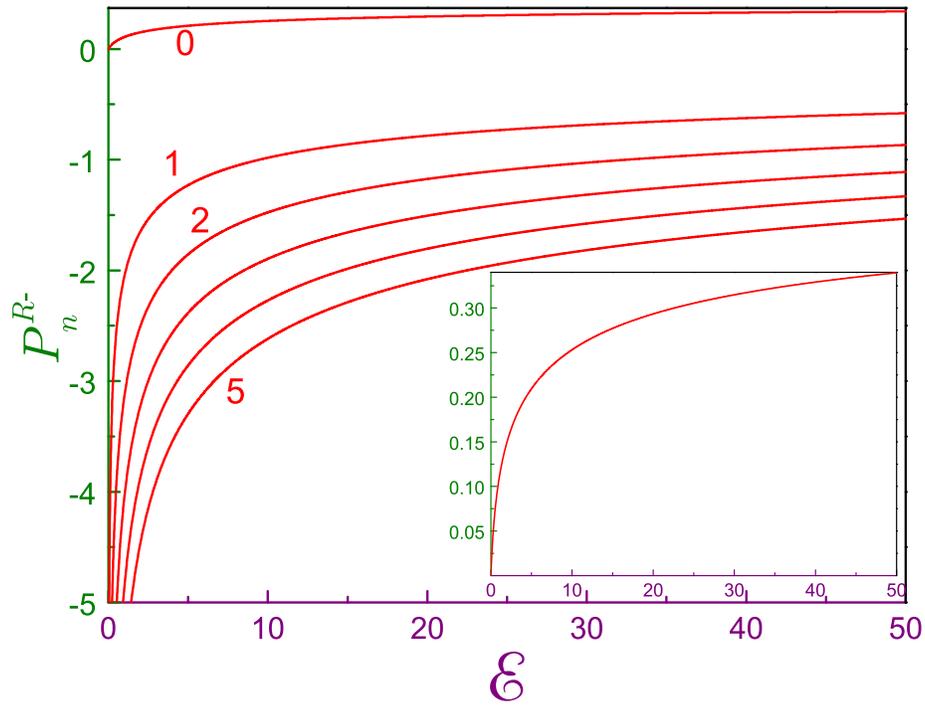}
\caption{\label{PolarizationFig1}
Dipole moments $P_n^{R-}$ of the negative Robin wall as function of the electric intensity $\mathscr{E}$. Numbers near the curves denote quantum numbers $n$. Inset shows enlarged view of the ground-state polarization.}
\end{figure}

Fig.~\ref{Functions} exhibits evolution with the electric intensity of the wave functions $\Psi_n(x)$ of the three lowest states for the negative extrapolation length where the level $n=0$ stays localized in the absence of the field while the higher lying states, $n=1,2,\ldots$, turn bound at $\mathscr{E}>0$ only. The increasing field monotonically pushes the waveform closer to the wall. Quantitative measure of the particle response to the applied voltage is provided by the dipole moment $P_n$. Calculation of polarization requires a knowledge of the zero-field mean value of the coordinate $x$, Eq.~\eqref{Polarization1}. For the ground state an elementary computation yields: $\langle x\rangle_{\mathscr{E}=0}=-1/2$ for $n=0$. Situation is different for the field-induced bound levels. Observe that in the limit of the vanishing electric intensities their energies tend to zero, see Eq.~\eqref{AsymptoticDirichlet1_NP}. But the nontrivial solution of the Schr\"{o}dinger equation with $E=\mathscr{E}=0$
\begin{equation}\label{FakeSolution1}
\Psi_{n\geq1}^{R\mp}(x)\sim x\pm1
\end{equation}
diverges at the negative infinity and should be dropped as an unphysical one. Accordingly, for the field-induced bound state a mean coordinate is zero:
\begin{equation}\label{PolarizationZero1}
\langle x\rangle_{\mathscr{E}=0}^{R-}=\left\{\begin{array}{cc}
-\frac{1}{2},&n=0\\
0,&n\geq1.
\end{array}\right.
\end{equation}
Plugging in the waveforms from Eqs.~\eqref{WaveFunction1_Robin} into the integral that defines the polarization, Eq.~\eqref{AngleBrackets1}, one is able to calculate it analytically \cite{Vallee1}:
\begin{subequations}\label{Polarization2}
\begin{eqnarray}\label{Polarization2_Minus}
P_n^{R-}(\mathscr{E})&=&-\frac{1}{3}\,\frac{2E_n^{R-}\left(E_n^{R-}+1\right)/\mathscr{E}+1}{E_n^{R-}+1}+\frac{1}{2}\delta_{n0}\\
\label{Polarization2_Plus}
P_n^{R+}(\mathscr{E})&=&-\frac{1}{3}\,\frac{2E_n^{R+}\left(E_n^{R+}+1\right)/\mathscr{E}-1}{E_n^{R+}+1}.
\end{eqnarray}
\end{subequations}
An alternative method applies Hellmann-Feynman theorem to the Hamiltonian from Eq.~\eqref{Hamiltonian2} \cite{Olendski3,Moyer1,Montgomery1}:
\begin{equation}\label{Hellmann1}
\langle x\rangle=-\left\langle\frac{\partial\hat{H}}{\partial\mathscr{E}}\right\rangle=-\frac{d E_n}{d\mathscr{E}}.
\end{equation}
Dropping the middle term in this equality chain and applying to Eq.~\eqref{EigenValueRobin1}, which for this purpose we will write in the form
\begin{equation}\label{EigenValueRobin2}
F(E,\mathscr{E})=0,
\end{equation}
the rule of differentiation of the implicit functions
\begin{equation}\label{Implicit1}
\frac{dE}{d\mathscr{E}}=-\frac{\partial F/\partial\mathscr{E}}{\partial F/\partial E},
\end{equation}
one arrives at Eqs.~\eqref{Polarization2}. Hellmann-Feynman theorem very easily provides expressions for the limiting cases of the strong and weak fields:
\begin{subequations}\label{PolarizationLimits1}
\begin{eqnarray}\label{PolarizationLimits1_SmallFields0}
P_0^{R-}(\mathscr{E})&=&\frac{1}{4}\mathscr{E},\quad\mathscr{E}\ll1\\
\label{PolarizationLimits1_SmallFieldsN}
P_n^{R-}(\mathscr{E})&=&\frac{2}{3}\frac{a_n}{\mathscr{E}^{1/3}},\quad n\geq1,\,\mathscr{E}\ll1\\
\label{PolarizationLimits1_LargeFieldsN}
P_n^{R-}(\mathscr{E})&=&\frac{2}{3}\frac{a_{n+1}'}{\mathscr{E}^{1/3}}+\frac{1}{2}\delta_{n0},\quad n=0,1,2,\ldots,\,\mathscr{E}\gg1.
\end{eqnarray}
\end{subequations}
It is seen that the ground-state polarization at the small electric intensities draws a straight line on the $\mathscr{E}-P$ plane when the linear contribution to the corresponding energy from Eq.~\eqref{AsymptoticDirichlet1_Negative0} exactly compensates the zero-field term. Observe that the dependence from Eq.~\eqref{PolarizationLimits1_SmallFields0} stays the same for the opposite direction of the field \cite{Olendski6,Moyer1}. Wave functions of the field-induced bound states in the same regime $\mathscr{E}\ll1$ simplify to
\begin{subequations}\label{AsymptoteFunction1}
\begin{align}\label{AsymptoteFunction1_Psi}
\Psi_n^{R-}(x)=-\frac{\mathscr{E}^{1/6}}{{\rm Ai}'\left(a_n\right)}{\rm Ai}\left(-\mathscr{E}^{1/3}x+a_n-\mathscr{E}^{1/3}\right),\,n\geq1,\\
\intertext{what means that its $m$th extremum, $m=1,2,\ldots,n$ is located at}
\label{AsymptoteFunction1_ExtremaLocation}
x_{nm}^{ext}=\frac{a_n-a_m'}{\mathscr{E}^{1/3}}-1,\, n\geq1,\,m=1,2,\ldots n,
\intertext{and the function values at these points are}
\label{AsymptoteFunction1_ExtremaValues}
\Psi_n^{R-}\!\left(x_{nm}^{ext}\right)=-\frac{{\rm Ai}(a_m')}{{\rm Ai}'(a_n)}\mathscr{E}^{1/6},\, n\geq1,\,m=1,2,\ldots n.
\intertext{Thus, the function maxima and minima for the decreasing weak field move away from the wall and the amplitude of its oscillations diminishes as $\mathscr{E}^{1/6}$. This results in the diverging polarization, as exemplified by Eq.~\eqref{PolarizationLimits1_SmallFieldsN} and Fig.~\ref{PolarizationFig1}, which shows several dipole moments of the negative Robin surface. Flattening of the waveforms as the intensity $\mathscr{E}$ fades is seen in Fig.~\ref{Functions}. For future reference, we also provide here the expression for the ground-state function at the weak fields, $\mathscr{E}\ll1$:}
\label{AsymptoteFunction1_Psi0}
\Psi_0^{R-}(x)=2^{1/2}\left[1+\frac{1}{4}\mathscr{E}\left(-x^2+\frac{1}{2}\right)\right]e^x.
\end{align}
\end{subequations}
Note that up to the first nonvanishing order of $\mathscr{E}$ it does satisfy the normalization, Eq.~\eqref{Normalization1}, and boundary, Eq.~\eqref{Robin3}, conditions and, when used in the integrals for calculating the polarization, Eq.~\eqref{AngleBrackets1}, yields again the linear dependence from Eq.~\eqref{PolarizationLimits1_SmallFields0}.

In the opposite limit of the high voltages, $\mathscr{E}\gg1$, one has 
\begin{subequations}\label{AsymptoteFunction2}
\begin{eqnarray}
\Psi_n^{R-}(x)&=&\frac{\mathscr{E}^{1/6}}{\left|-a_{n+1}'\right|^{1/2}{\rm Ai}\!\left(a_{n+1}'\right)}\nonumber\\
\label{AsymptoteFunction2_Psi}
&\times &{\rm Ai}\!\left(-\mathscr{E}^{1/3}x+a_{n+1}'+\frac{1}{a_{n+1}'\mathscr{E}^{1/3}}\right)\\
\label{AsymptoteFunction2_ExtremaLocation}
x_{nm}^{ext}&=&\frac{a_{n+1}'-a_{m+1}'}{\mathscr{E}^{1/3}}+\frac{1}{a_{n+1}'\mathscr{E}^{2/3}}\\
\label{AsymptoteFunction2_ExtremaValues}
\Psi_n^{R-}\!\left(x_{nm}^{ext}\right)&=&\frac{{\rm Ai}\left(a_{m+1}'\right)}{{\rm Ai}\left(a_{n+1}'\right)}\frac{\mathscr{E}^{1/6}}{\left|-a_{n+1}'\right|^{1/2}},
\end{eqnarray}
\end{subequations}
$n=0,1,\ldots$, $m=0,1,\ldots n$. It means that at the strong electric intensities the particle gets squeezed to the wall what has it consequence in the decreasing of the absolute value of the dipole moment down to zero. Since the power of the electric field entering Eq.~\eqref{PolarizationLimits1_LargeFieldsN} is quite small, the approach to this limit is relatively slow, as Fig.~\ref{PolarizationFig1} demonstrates. The buildup of the electron density close to the Robin surface with the growing voltage is also exemplified in Fig.~\ref{Functions}.

Nondiagonal dipole moment matrix elements are calculated as
\begin{subequations}\label{PolarizMatrix2}
\begin{eqnarray}\label{PolarizMatrix2_R}
P_{nm}^R&=&\frac{\mathscr{E}}{\left[\left(E_n^R+1\right)\left(E_m^R+1\right)\right]^{1/2}}\frac{E_n^R+E_m^R+2}{\left(E_n^R-E_m^R\right)^2}\\
\label{PolarizMatrix2_D}
P_{nm}^D&=&\frac{2}{\left(a_{n+1}-a_{m+1}\right)^2}\,\mathscr{E}^{-1/3}\\
\label{PolarizMatrix2_N}
P_{nm}^N&=&-\frac{1}{\left(a_{n+1}'a_{m+1}'\right)^{1/2}}\frac{a_{n+1}'+a_{m+1}'}{\left(a_{n+1}'-a_{m+1}'\right)^2}\,\mathscr{E}^{-1/3}.
\end{eqnarray}
\end{subequations}
Observe that in the limiting cases the Robin matrix elements transform into their Dirichlet or Neumann counterparts:
\begin{equation}\label{PolarizMatrix3}
P_{nm}^R=\left\{\begin{array}{cc}
P_{nm}^D,&\mathscr{E}\ll1\\
P_{nm}^N,&\mathscr{E}\gg1.
\end{array}
\right.
\end{equation}
Special care has to be taken for the transitions involving the ground level of the negative Robin wall at the low voltages, $\mathscr{E}\ll1$:
\begin{equation}\label{PolarizMatrix4}
P_{0n}^{R-}=\left\{\begin{array}{cc}
2^{1/2}\mathscr{E}^{1/2},&|a_{n+1}|\mathscr{E}^{2/3}\ll1\\
\left(-2\!\left/a_{n+1}^3\right.\right)^{1/2}\mathscr{E}^{-1/2},&|a_{n+1}|\mathscr{E}^{2/3}\gg1.
\end{array}
\right.
\end{equation}
Eq.~\eqref{PolarizMatrix4} shows that in this regime the matrix element dependence on the field changes from the square root for the low lying states to its inverse for the very large $n$.

\section{Quantum Information Measures}\label{Sec_QuantumInfo}
\subsection{Quantum Information Entropy}\label{Sec_Entropy}

\begin{figure*}
\centering
\includegraphics*[width=\columnwidth]{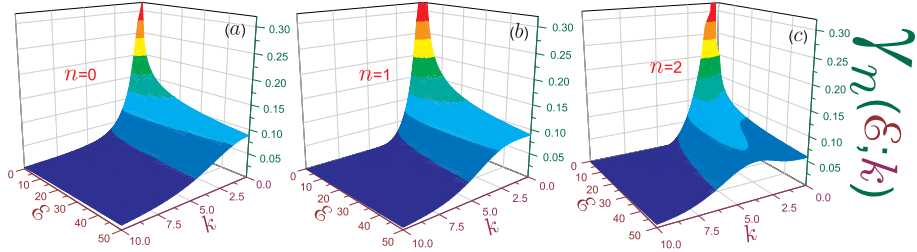}
\caption{\label{MomentumDensity}
Momentum density $\gamma_n(k)$ of the (a) lowest level, (b) first and (c) second field-induced bound states of the negative Robin wall in terms of the momentum $k$ and electric intensity $\mathscr{E}$.}
\end{figure*}

Knowledge of the energies $E_n$ and functions $\Psi_n(x)$ allows calculation of other physical parameters; for example,  dipole moment  as a function of the field was calculated in the previous section. Position Shannon entropy $S_x$ is computed directly from Eq.~\eqref{EntropyX_1} with the use of Eq.~\eqref{densityX_1} while for finding quantum information entropy in the momentum space $S_k$, Eq.~\eqref{EntropyK_1}, one needs to evaluate first the momentum wave function $\Phi_n(k)$ according to Eq.~\eqref{FunctionPhi1}. In known to me literature \cite{Abramowitz1,Vallee1,Gradshteyn1,Prudnikov2,Prudnikov3} there are no analytical expressions for the integrals in Eqs.~\eqref{FunctionPhi1}, \eqref{EntropyX_1} and \eqref{EntropyK_1} with the functions from Eqs.~\eqref{WaveFunction1}; accordingly, their direct numerical quadrature was performed in calculating the results presented below. Fig. \ref{MomentumDensity} depicts a transformation of the momentum density $\gamma_n(k)$ of the several low lying states of the negative Robin length as the electric field varies. As $\gamma_n(k)$ is an even function of its argument, $\gamma_n(-k)=\gamma_n(k)$, the parts with the non negative $k$ only are shown. For the field-free case, the functions are:
\begin{equation}\label{FunctionPhi3}
\gamma_n(k)=\left\{
\begin{array}{cc}\frac{1}{\pi}\frac{1}{1+k^2},&n=0\\
\delta(k),&n\ne0.
\end{array}
\right.
\end{equation}
Characteristic feature of the field influence is the decrease of the maximum value at $k=0$ (finite, $1/\pi$ for $n=0$ and infinite for $n\geq1$) as the applied voltage grows. For the ground state using the position waveform from Eq.~\eqref{AsymptoteFunction1_Psi0} one finds:
\begin{subequations}\label{AsymptoteFunction3}
\begin{align}\label{AsymptoteFunction3_Psi0}
\Phi_0^{R-}(k)&=\frac{1}{\pi^{1/2}}\frac{1}{1-ik}\left\{1+\frac{\mathscr{E}}{4}\left[1-\frac{2}{(1-ik)^2}\right]\right\},\,\mathscr{E}\ll1,
\intertext{and, accordingly:}
\label{AsymptoteFunction3_gamma0}
\gamma_0^{R-}(k)&=\frac{1}{\pi}\frac{1}{1+k^2}\left[1+\mathscr{E}\left(\frac{1}{2}-\frac{1}{1+k^2}\right)\right],\quad\mathscr{E}\ll1,
\intertext{what means that the ground-state zero momentum, which remains the only maximum of the function $\gamma_0(k)$ for all fields, linearly decreases with the growth of the small $\mathscr{E}$:}
\label{AsymptoteFunction3_gamma00}
{\gamma_0^{R-}}_{\!\!\!\!max}&\equiv\gamma_0^{R-}(0)=\frac{1}{\pi}\left(1-\frac{1}{2}\mathscr{E}\right),\quad\mathscr{E}\ll1.
\end{align}
\end{subequations}
As Fig.~\ref{MomentumDensity} shows, for the higher lying levels the two symmetric with respect to $k=0$ peaks are formed with the field growing. The distance between these gentle maxima increases with the quantum number $n$.

\begin{figure}
\centering
\includegraphics*[width=\columnwidth]{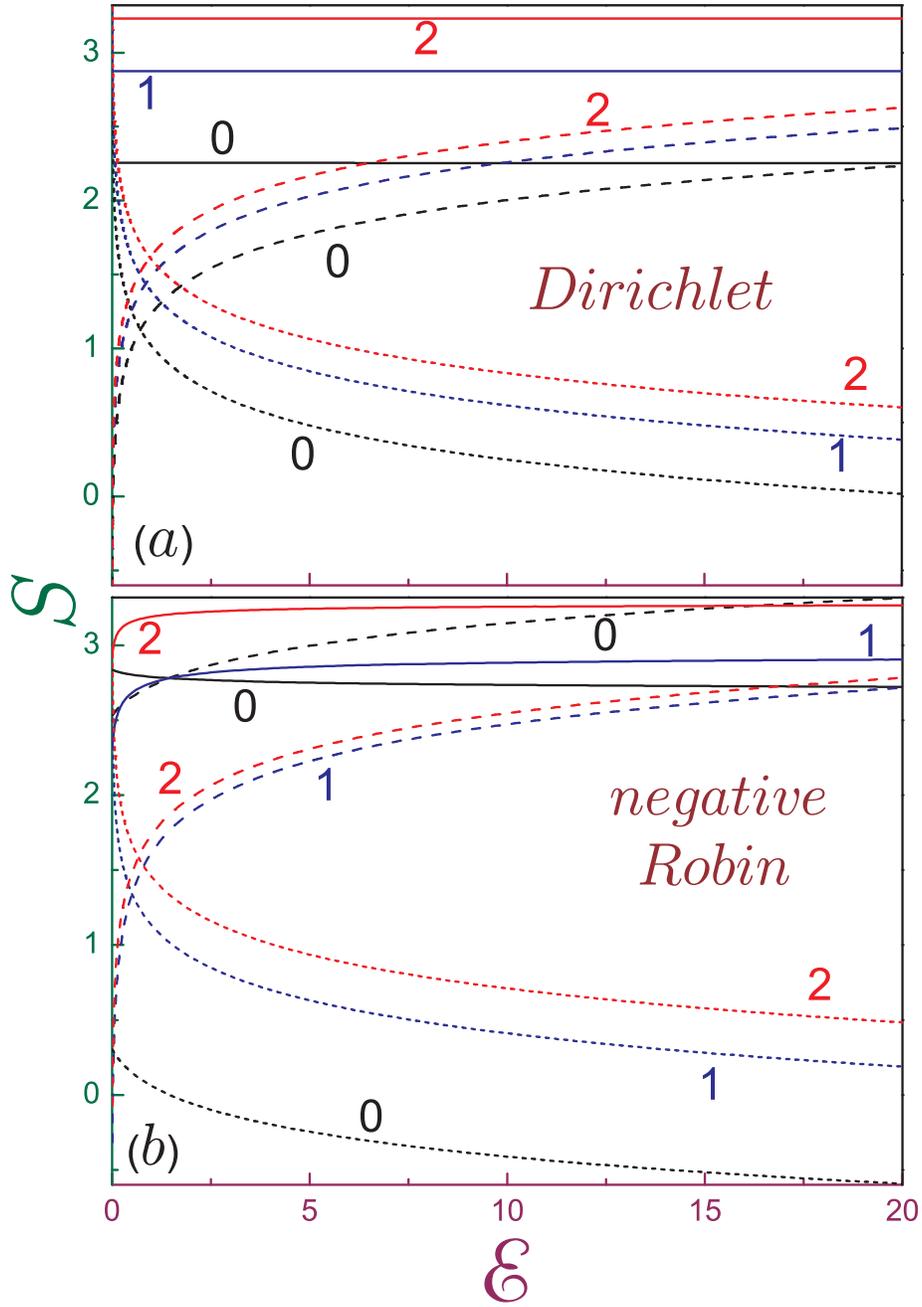}
\caption{\label{Entropies}
Position $S_x$ (dotted lines), momentum $S_k$ (dashed lines) and total $S_t$ (solid curves) entropies as a function of the electric field $\mathscr{E}$ for (a) Dirichlet and (b) negative Robin wall. The digits near the curves denote the corresponding quantum number $n$.}
\end{figure}

Fig.~\ref{Entropies} exhibits position $S_x$, momentum $S_k$ entropies and their sum $S_t$ of the three lowest levels as functions of the electric intensity for the Dirichlet [panel (a)] and negative Robin wall [panel (b)]. Considering the Dirichlet case, we remind again that its all bound states are formed as a result of the applied voltage. It is elementary to show that their position ${S_x^D}_{\!\!n}$ and momentum ${S_k^D}_{\!\!n}$ entropies depend logarithmically on the field with, however, the opposite signs:
\begin{subequations}\label{DirichletEntropy1}
\begin{eqnarray}
&&{S_x^D}_{\!\!n}=-\frac{1}{3}\ln\mathscr{E}+\ln\!\left({\rm Ai}'\!\left(a_{n+1}\right)^2\right)\nonumber\\
\label{DirichletEntropyX1}
&-&\frac{1}{{\rm Ai}'\!\left(a_{n+1}\right)^2}\int_{-\infty}^0\!\!\!\!\!{\rm Ai}^2(-x+a_{n+1})\ln{\rm Ai}^2(-x+a_{n+1})\,dx\\
&&{S_k^D}_{\!\!n}=\frac{1}{3}\ln\mathscr{E}+\ln\!\left(2\pi{\rm Ai}'\!\left(a_{n+1}\right)^2\right)\nonumber\\
&-&\frac{1}{2\pi{\rm Ai}'\!\left(a_{n+1}\right)^2}\int_{-\infty}^\infty\left|\int_{-\infty}^0e^{ikx}{\rm Ai}(-x+a_{n+1})dx\right|^2\nonumber\\
\label{DirichletEntropyK1}
&\times&\ln\left|\int_{-\infty}^0e^{ikx}{\rm Ai}(-x+a_{n+1})dx\right|^2dk.
\end{eqnarray}
\end{subequations}
Accordingly, their sum $S_t^D=S_x^D+S_k^D$ is the field independent quantity, as Fig.~\ref{Entropies} demonstrates where also the logarithmic dependencies of the entropies ${S_x^D}_{\!\!n}$ and ${S_k^D}_{\!\!n}$ are clearly seen. It is very instructive to compare exact results  from Eqs.~\eqref{DirichletEntropy1} with the approximate method that replaces the triangular potential by the infinitely deep {\em flat} Dirichlet quantum well with the field- and level-dependent width $L_{ap}$; for example, for the lowest Dirichlet level, $n=0$, utilizing the form of the Airy function \cite{Abramowitz1,Vallee1}, it is reasonable to assume that
\begin{equation}\label{Lapprox1}
L_{ap}=\frac{2|a_1|}{\mathscr{E}^{1/3}}
\end{equation}
what means that the motion of the particle is governed not by the function $\Psi_0^D(x)$ from Eq.~\eqref{WaveFunction1_Dirichlet} but by
\begin{equation}\label{FunctionApproximate1}
\Psi_{ap}(x)=\left(\frac{2}{L_{ap}}\right)^{1/2}\sin\frac{\pi}{L_{ap}}\,x,\quad-L_{ap}\le x\le0.
\end{equation}
Calculating corresponding position and momentum entropies produces \cite{Olendski3}:
\begin{subequations}\label{ApproximateEntropies1}
\begin{eqnarray}
S_x^{ap}&=&2\ln2-1+\ln|a_1|-\frac{1}{3}\,\mathscr{E}\nonumber\\
\label{ApproximateEntropiesX1}
&=&S_{x_0}^{(1)}+\ln2+\ln|a_1|-\frac{1}{3}\ln\mathscr{E}\\
\label{ApproximateEntropiesK1}
S_k^{ap}&=&S_{k_0}^{(1)}-\ln2-\ln|a_1|+\frac{1}{3}\ln\mathscr{E},
\end{eqnarray}
\end{subequations}
where $S_{x_0}^{(1)}=\ln2-1\approx-0.30685$ and $S_{k_0}^{(1)}\approx2.5189$ being corresponding entropies of the unit-length Dirichlet quantum well \cite{Olendski3}. Eqs.~\eqref{ApproximateEntropies1} prove again the logarithmic dependence of the Dirichlet position and momentum entropies on the field and immediately yield:
\begin{equation}\label{ApproximateEntropies2}
S_t^{ap}=S_{x_0}^{(1)}+S_{k_0}^{(1)}=S_{t_0}^{(1)}\approx2.212,
\end{equation}
what, of course, satisfies entropic uncertainty relation, Eq.~\eqref{EntropicInequality1}, for $1+\ln\pi\approx2.145$ \cite{Olendski3}. Thus, the total ground-state entropy in this approximation is equal to its counterpart of the flat unit-length Dirichlet quantum well. Exact calculations show that ${S_t^D}_{\!\!0}=2.254$. It is quite remarkable that such a crude approximation as the one used above provides such a good accuracy. It should be noted however that for the higher-lying states the entropy $S_t^D$ diverges from the corresponding values of the field-free well.

In the same way, it can be shown that for the Neumann BC the position and momentum entropies depend on the field as a one third of $\ln\mathscr{E}$ and that their sum is independent of the applied voltage too. Situation changes for the Robin BC; for example, for the negative wall the field-induced levels are, as stated above, Dirichlet-like at the small electric forces, Eq.~\eqref{AsymptoticDirichlet1_NP}, and Neumann ones at the large intensities, Eq.~\eqref{AsymptoticNeumann1}. Hence, their dependence on the field in these asymptotic cases is logarithmic too; however, the coefficients describing entropies for $\mathscr{E}\gg1$ are different from those given above. As a result of the transition from one type of the BC to another one with varying electric force, the total entropies of these states are field-dependent, as panel (b) demonstrates.  The field-free lowest state possesses finite entropies $S_x=1-\ln2$ and $S_k=2\ln2+\ln\pi$, see Eqs.~\eqref{SingleRobinWallEntropy}. For the small fields, utilizing asymptotic forms of the functions from Eqs.~\eqref{AsymptoteFunction1_Psi0} and \eqref{AsymptoteFunction3_gamma0}, one finds:
\begin{subequations}\label{AsymptoteEntropy1}
\begin{eqnarray}\label{AsymptoteEntropy1_X}
{S_x^{R-}}_{\!\!\!\!\!0}(\mathscr{E})=1-\ln2-\frac{1}{2}\mathscr{E},\quad\mathscr{E}\ll1\\
\label{AsymptoteEntropy1_K}
{S_k^{R-}}_{\!\!\!\!\!0}(\mathscr{E})=\ln(4\pi)+\frac{1}{2}\mathscr{E},\quad\mathscr{E}\ll1,
\end{eqnarray}
\end{subequations}
what means that the sum of the two entropies in this regime depends on the electric intensity as the power higher than the linear one. With the voltage varying the ground level undergoes a transformation from the Robin BC at the small fields to the Neumann one at the large $\mathscr{E}$ what results again in the field-dependent sum $S_t$. Larger magnitude of the negative position entropy at the high electric intensities means stronger localization of the electron. Mathematically, it is obvious why for the ground state it is in this limit greater for the Neumann or Robin BC than for the Dirichlet requirement
\begin{equation}\label{EntropicInequality2}
\left|{S_x^{R,N}}_{\!\!\!\!\!\!\!\!0}\,\,\,\right|>\left|{S_x^D}_{\!\!0}\right|,\quad\mathscr{E}\gg1;
\end{equation}
namely, for the former case the maximum of the function with its magnitude larger than unity, see panel (a) in Fig.~\ref{Functions}, adds a large negative contribution to the position entropy $S_x$ from Eq.~\eqref{EntropyX_1} while for the Dirichlet wall it is partially compensated by the fraction of the function $\Psi_0^D(x)$ that lies in the very vicinity of the interface with its value smaller than one. Since higher lying states have several extrema with the alternating signs, the difference between positions entropies for the different BCs decreases, as it is seen from Fig.~\ref{Entropies}. In the same way, zero-field value of the momentum entropy of the lowest Robin state is positive since the maximum of the momentum density is smaller than unity, see Eq.~\eqref{FunctionPhi3}. Growing electric field decreases the maximum of $\gamma_0^{R-}(k)$ as exemplified by Eq.~\eqref{AsymptoteFunction3_gamma00} and shown in panel (a) of Fig.~\ref{MomentumDensity} what physically means a larger uncertainty in determining the momentum and leads to the increase of the corresponding entropy from Eq.~\eqref{EntropyK_1}. It is also important to note that the previously discovered rule \cite{Sun1,Laguna1,Olendski3} of the increase of the total entropy $S_t$ with the quantum number $n$, in general, does not hold here; namely, as panel (b) of Fig.~\ref{Entropies} shows, there is a range of the fields $0<\mathscr{E}<\mathscr{E}_\times$, where 
\begin{equation}\label{EntropicInequality3}
{S_t^{R-}}_{\!\!\!\!\!\!0}\,>{S_t^{R-}}_{\!\!\!\!\!\!1}\quad{\rm at}\quad0<\mathscr{E}<\mathscr{E}_\times.
\end{equation}
This is explained by the fact that at the small voltages these two levels obey different BCs: the lowest (upper lying) state is the Robin (Dirichlet) -like one, as it was stressed above; accordingly, different edge requirements determine different entropies in such a way that Eq.~\eqref{EntropicInequality3} holds. With the growth of the field, all the levels of the negative Robin wall tend to become the Neumann ones; as a result, their entropies rearrange in a regular order when the sum $S_t$ is larger for the higher lying states. Exact calculation shows that $\mathscr{E}_\times=1.45$.

\subsection{Fisher Information}\label{Sec_Fisher}
\begin{figure}
\centering
\includegraphics*[width=\columnwidth]{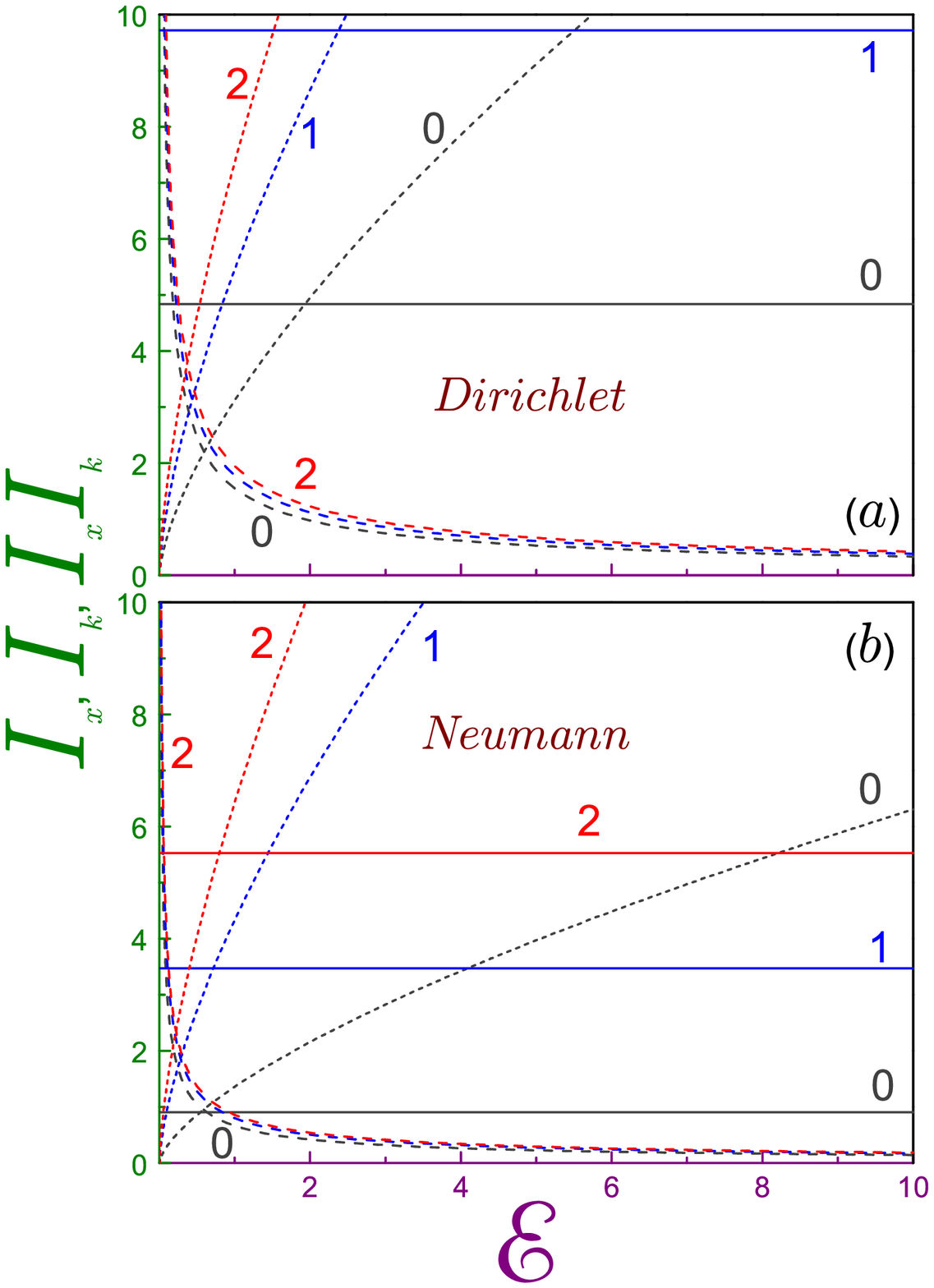}
\caption{\label{FisherDirichletNeumannFig1}
Fisher position $I_x$ (dotted lines), momentum $I_k$ (dashed lines) informations and their product $I_xI_k$ (solid curves) as a function of the electric field $\mathscr{E}$ for (a) Dirichlet and (b) Neumann wall. The digits near the curves denote the corresponding quantum number $n$.}
\end{figure}

Position Fisher information $I_x$ in the electric field is calculated analytically as
\begin{subequations}\label{FisherX2}
\begin{eqnarray}\label{FisherX2_R}
I_{x_n}^R&=&\frac{4}{3}\frac{2\mathscr{E}+E_n+E_n^2}{E_n+1}\\
\label{FisherX2_DN}
I_{x_n}^{D,N}&=&\frac{4}{3}E_n^{D,N}.
\end{eqnarray}
\end{subequations}
So, the Dirichlet or Neumann position Fisher information is up to the factor $4/3$ simply the corresponding energy of the state what means that they vary as $\mathscr{E}^{2/3}$. For the Robin surface in the limits of the weak and strong voltages it reduces to the Dirichlet or Neumann case, respectively:
\begin{subequations}\label{FisherX3}
\begin{align}\label{FisherX3_N}
I_{x_n}^{R\mp}&=\left\{\begin{array}{cc}
\left\{\begin{array}{c}I_{x_{n-1}}^D\\I_{x_n}^D\end{array}\right\},&\mathscr{E}\ll1\\
I_{x_n}^N,&\mathscr{E}\gg1.
\end{array}\right.
\intertext{The only exception is the ground level of the attractive Robin wall at the vanishing electric intensities:}
\label{FisherX3_0}
I_{x_0}^{R-}&=4+2\mathscr{E}+\frac{1}{6}\mathscr{E}^2,\quad\mathscr{E}\ll1.
\end{align}
\end{subequations}
\begin{figure*}
\centering
\includegraphics*[width=1.1\columnwidth]{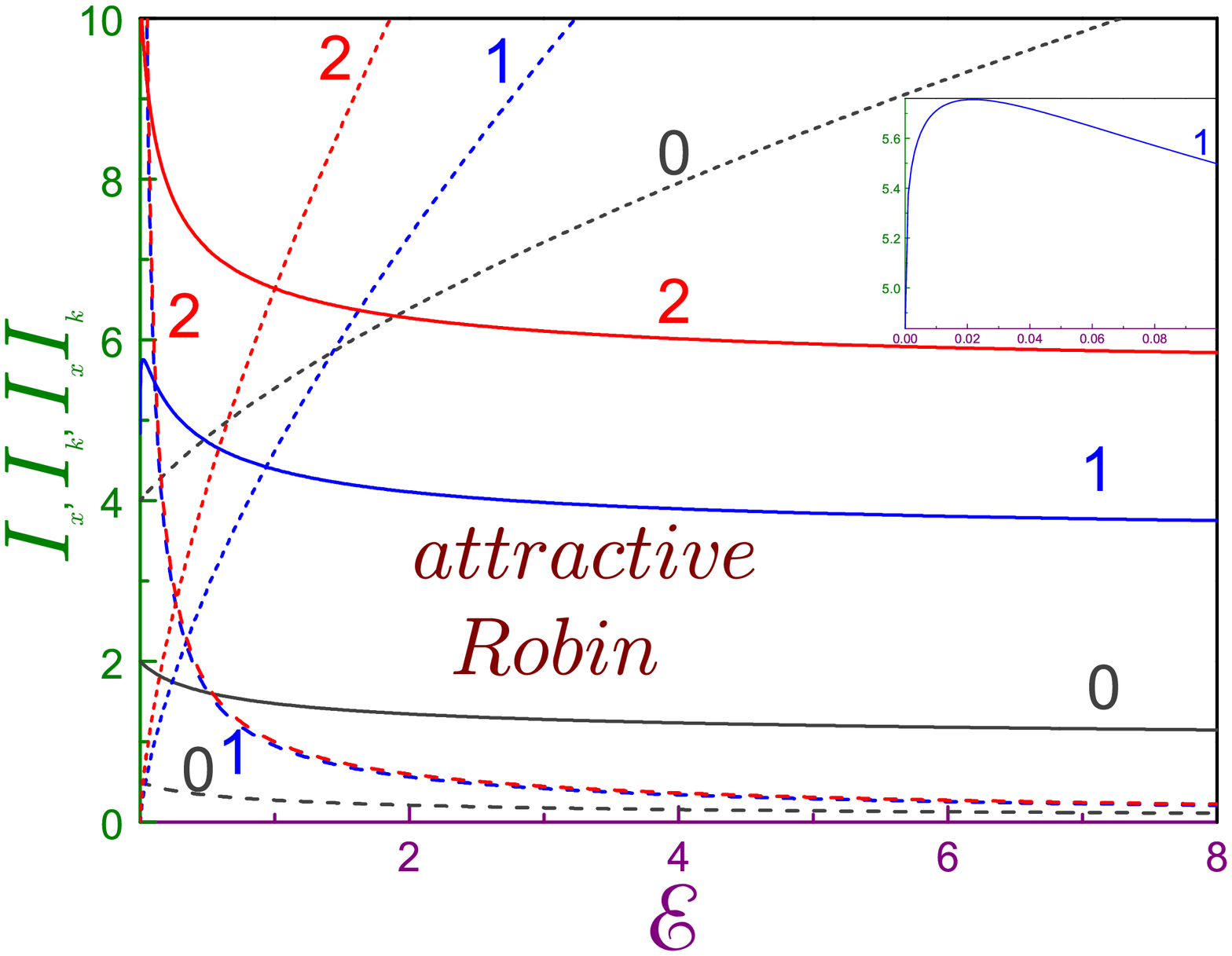}
\caption{\label{FisherRobinFig1}
The same as in Fig.~\ref{FisherDirichletNeumannFig1} but for the attractive Robin wall. Inset shows an enlarged view of the product $I_xI_k$ of the first excited state at the small fields.}
\end{figure*}

Expressions for the Dirichlet and Neumann momentum Fisher informations can be written as:
\begin{equation}\label{FisherK1}
I_{k_n}^{D,N}=\frac{C_n^{D,N}}{\mathscr{E}^{2/3}},
\end{equation}
where the field-independent coefficients $C_n^{D,N}$ are expressed with the help of the integrals whose integrands contain the Airy functions. Due to unwieldiness, we do not write their explicit forms here. Eqs.~\eqref{FisherX2_DN} and \eqref{FisherK1} mean that the applied voltage can not change the product of the position and momentum Fisher informations $I_{x_n}I_{k_n}$ for the Dirichlet or Neumann BCs since the change with the field of the either factor in the product is exactly compensated by the opposite variation of the second multiplier. This conclusion is similar to the one about the sum of the entropies for the same edge requirements derived in the previous subsection.

Fig.~\ref{FisherDirichletNeumannFig1} shows the features of the Dirichlet and Neumann Fisher informations discussed above. Contrary to the quantum information entropies that decrease (increase) with the growing field as its natural logarithm, the position (momentum) Fisher information enlarges (diminishes) with the voltage proportionally to $\mathscr{E}^{2/3}$. Product of the two informations does not depend on the field, is larger for the larger quantum number and,  at the fixed $n$, is greater for the Dirichlet wall. Small position Fisher informations at the disappearing electric intensities are explained by the almost flat corresponding waveforms what makes their derivatives in Eq.~\eqref{Fisher1D_X1} very close to zero. Simultaneously, the diverging momentum Fisher informations are due to the very sharp shape of $|\Phi(k)|^2$, which are centered around $k=0$ and tend to the $\delta$-function at $\mathscr{E}\rightarrow0$. At the high voltages, the components exchange their roles: strong localization of $\Psi(x)$ at the interface implies the huge magnitude of its derivative leading to the unrestrictedly increasing $I_x$ while the vanishing speed of variation of the momentum function almost zeroes its part of the Fisher information.

Fig.~\ref{FisherRobinFig1} depicts Fisher informations for the attractive Robin surface. At the small fields, the ground-state position information increases from its zero-field value as described by Eq.~\eqref{FisherX3_0} while its momentum counterpart gets smaller as
\begin{subequations}\label{FisherK3}
\begin{align}\label{FisherK3_0}
I_{k_0}^{R-}=\frac{1}{2}(1-\mathscr{E}),\quad\mathscr{E}\ll1,
\intertext{what means that their product is a linearly decreasing function of the weak electric force:}
\label{FisherXK3_0}
I_{x_0}^{R-}I_{k_0}^{R-}=2-\mathscr{E},\quad\mathscr{E}\ll1.
\end{align}
\end{subequations}
At the high voltages they tend asymptotically to the values that correspond to the lowest Neumann level with the product $I_{x_0}^{R-}I_{k_0}^{R-}$ becoming less and less dependent on the electric intensity with the growth of the latter. A characteristic feature of the Fisher information of the field-induced bound states is a nonmonotonic dependence on $\mathscr{E}$ of the product $I_{x_n}^{R-}I_{k_n}^{R-}$, $n\geq1$, while its each multiplier is a constantly increasing ($I_{x_n}^{R-}$) or decreasing ($I_{k_n}^{R-}$) function of the applied voltage. For example, for the first excited level this product from its zero-field value of $4.837$ corresponding to the lowest Dirichlet state rapidly grows to the maximum of $5.756$ achieved at $\mathscr{E}=0.022$ after which it gently decreases approaching at the high electric intensities the value of $3.472$ of its Neumann counterpart. A detailed form of this maximum, which is also characteristic for the higher lying states too, is shown in the inset.

\subsection{Onicescu Information Energy and complexity CGL}\label{Sec_Onicescu}
\begin{figure}
\centering
\includegraphics*[width=\columnwidth]{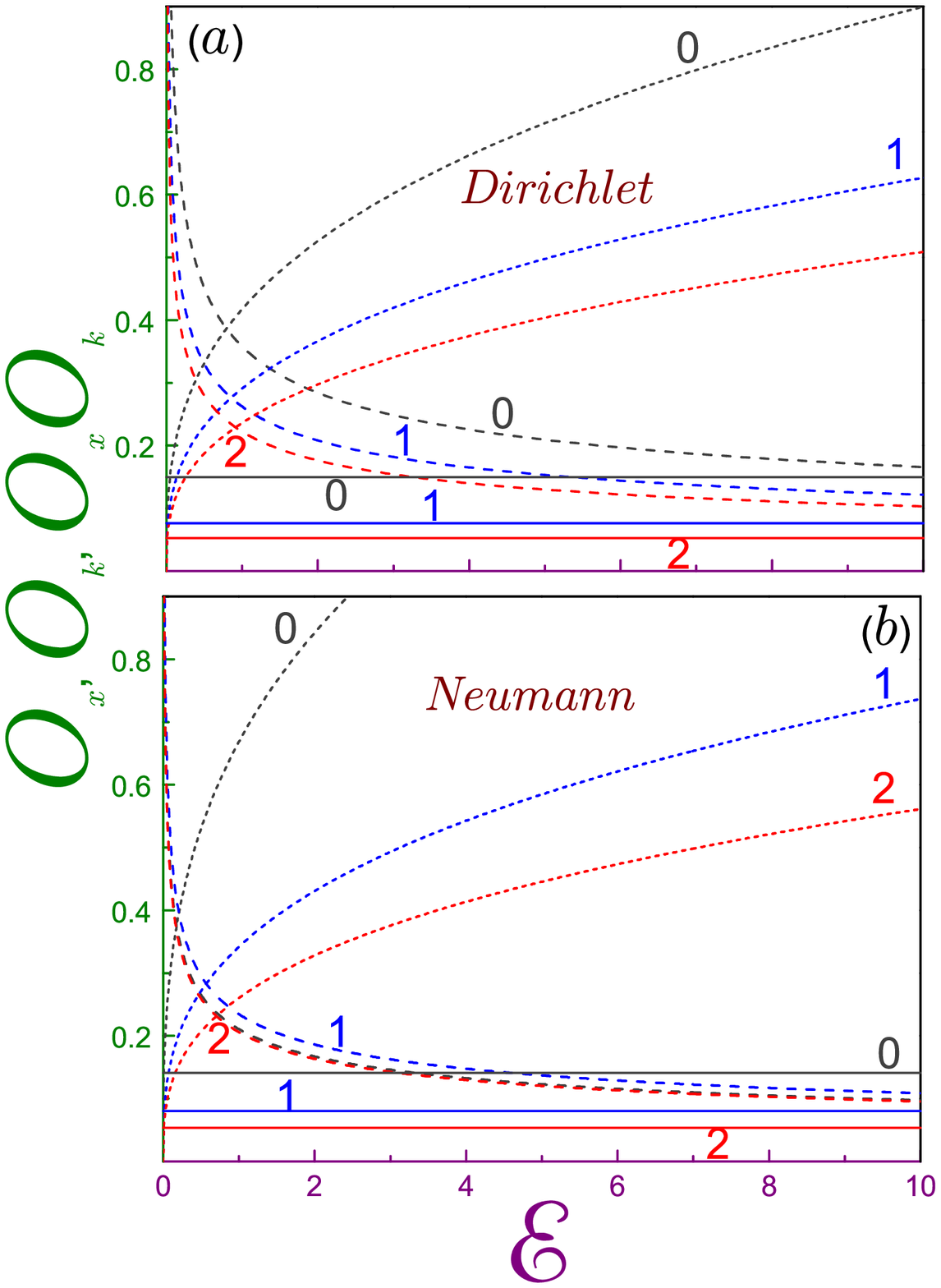}
\caption{\label{OnicescuDirichletNeumannFig1}
Onicescu position $O_x$ (dotted curves), momentum $O_k$ (dashed curves) information energies and their product $O_xO_k$ (solid lines) as a function of the electric field $\mathscr{E}$ for (a) Dirichlet and (b) Neumann wall. The digits near the curves denote the corresponding quantum number $n$.}
\end{figure}

Dirichlet and Neumann position information energies are proportional to the cubic root of the field while their momentum counterparts are inverse functions of it:
\begin{subequations}\label{OnicescuDN1}
\begin{eqnarray}\label{OnicescuDN1_Dx}
O_{x_n}^D&=&\frac{\mathscr{E}^{1/3}}{{\rm Ai}'\left(a_{n+1}\right)^4}\int_0^\infty{\rm Ai}^4\left(x+a_{n+1}\right)dx\\
\label{OnicescuDN1_Nx}
O_{x_n}^N&=&\frac{\mathscr{E}^{1/3}}{\left|a_{n+1}'\right|^2{\rm Ai}^4\left(a_{n+1}'\right)}\int_0^\infty{\rm Ai}^4\left(x+a_{n+1}'\right)dx\\
O_{k_n}^D&=&\frac{1}{(2\pi)^2}\frac{\mathscr{E}^{-1/3}}{{\rm Ai}'\left(a_{n+1}\right)^4}\nonumber\\
\label{OnicescuDN1_Dk}
&\times&\int_{-\infty}^\infty dk\left|\int_0^\infty dxe^{ikx}{\rm Ai}\left(x+a_{n+1}\right)\right|^4\\
O_{k_n}^N&=&\frac{1}{(2\pi)^2}\frac{\mathscr{E}^{-1/3}}{\left|a_{n+1}'\right|^2{\rm Ai}\left(a_{n+1}'\right)^4}\nonumber\\
\label{OnicescuDN1_Nk}
&\times&\int_{-\infty}^\infty dk\left|\int_0^\infty dxe^{ikx}{\rm Ai}\left(x+a_{n+1}'\right)\right|^4.
\end{eqnarray}
\end{subequations}
Thus, similar to the Fisher information for the same BCs, the product of the two Onicescu energies is a field-independent quantity.  Numerical evaluation of the integrals from Eqs.~\eqref{OnicescuDN1} is shown in Fig.~\ref{OnicescuDirichletNeumannFig1}. Observe that, contrary to the position and momentum entropies and their sum and to the Fisher informations and their products, the Onicescu energies are decreasing functions of the quantum number $n$:
\begin{subequations}\label{OnicescuInequalities1}
\begin{eqnarray}\label{OnicescuInequalities1_X}
O_{x_n}^{D,N}&>&O_{x_{n+1}}^{D,N},\quad n=0,1,2,\ldots\\
\label{OnicescuInequalities1_K}
O_{k_n}^{\left\{_N^D\right\}}&>&O_{k_{n+1}}^{\left\{_N^D\right\}},\quad\left\{\begin{array}{c}
n=0,1,2,\ldots\\
n=1,2,3\ldots
\end{array}\right.\\
\label{OnicescuInequalities1_XK}
O_{x_n}^{\left\{_N^D\right\}}\cdot O_{k_n}^{\left\{_N^D\right\}}&>&O_{x_{n+1}}^{\left\{_N^D\right\}}\cdot O_{k_{n+1}}^{\left\{_N^D\right\}},\quad n=0,1,2,\ldots
\end{eqnarray}
\end{subequations}
Moreover, the monotonic dependence of the quantum-information measures on the number $n$ is broken here since, for example, the Neumann momentum information energy for the lowest level at the arbitrary field lies in between its $n=1$ and $n=2$ counterparts: $O_{k_2}^N<O_{k_0}^N<O_{k_1}^N$. Nevertheless, Eq.~\eqref{OnicescuInequalities1_XK} holds for these states too. A comparison of the Dirichlet and Neumann BCs yields:
\begin{subequations}
\begin{eqnarray}
O_{x_n}^D&<&O_{x_n}^N\\
O_{k_n}^D&>&O_{k_n}^N\\
O_{x_n}^D\cdot O_{k_n}^D&\gtrless &O_{x_n}^N\cdot O_{k_n}^N,\quad\left\{\begin{array}{l}
n=0\\
n=1,2,\ldots.
\end{array}\right.
\end{eqnarray}
\end{subequations}

\begin{figure}
\centering
\includegraphics*[width=\columnwidth]{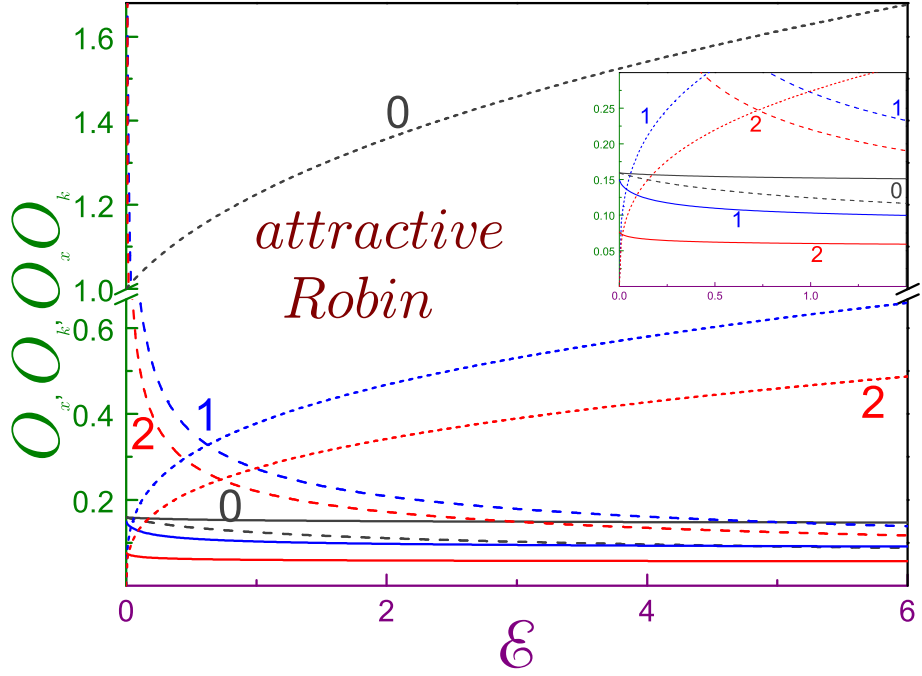}
\caption{\label{OnicescuRobinFig1}
The same as in Fig.~\ref{OnicescuDirichletNeumannFig1} but for the attractive Robin wall. Note vertical line break from $0.67$ to unity. Inset shows the enlarged view at the small electric fields.}
\end{figure}

Onicescu energies of the attractive Robin wall are shown in Fig.~\ref{OnicescuRobinFig1}. Zero-field information energies with the growing electric intensity move in the opposite directions: while the position one increases, its momentum counterpart decreases in such a way that the product of the two changes only slightly in a wide range of the voltages; for example, $O_{x_0}^{R-}O_{k_0}^{R-}$ from its zero-field value of $0.15915$ from Eq.~\eqref{Onicescu2_XK} drops down to $0.14395$ at $\mathscr{E}=40$, almost corresponding to the lowest Neumann level with $0.14081$. At the low voltages, $\mathscr{E}\ll1$, their dependencies are expressed analytically as
\begin{subequations}\label{OnicescuAsymptote1}
\begin{eqnarray}\label{OnicescuAsymptote1_X}
O_{x_0}^{R-}&=&1+\frac{3}{8}\mathscr{E}\\
\label{OnicescuAsymptote1_K}
O_{k_0}^{R-}&=&\frac{1}{2\pi}\left(1-\frac{1}{2}\mathscr{E}\right)\\
\label{OnicescuAsymptote1_XK}
O_{x_0}^{R-}O_{k_0}^{R-}&=&\frac{1}{2\pi}\left(1-\frac{1}{8}\mathscr{E}\right).
\end{eqnarray}
\end{subequations}

Information energies of the Robin field-induced states change their values corresponding to the Dirichlet BC at the low intensities $\mathscr{E}$ to the Neumann magnitudes at the high voltages. For these edge requirements, the product of the position and momentum energies is a monotonically decreasing function of the quantum number too.

\begin{table*}[ht]
\caption{Position $CGL_x$, momentum $CGL_k$ and their product $CGL_x\cdot CGL_k$ complexity measures for the Dirichlet and Neumann quantum walls and several low lying states}
\centering 
\begin{tabular}{c|c|c|c||c|c|c}
\hline\hline
\multirow{2}[3]{*}{$n$}&\multicolumn{3}{c||}{Dirichlet}&\multicolumn{3}{|c}{Neumann}\\[0.5ex]
\cline{2-4}\cline{5-7}
&$\begin{array}{c}{\rm Position}\\CGL_x\end{array}$&$\begin{array}{c}{\rm Momentum}\\CGL_k\end{array}$&$\begin{array}{c}{\rm Product}\\CGL_x\cdot CGL_k\end{array}$&$\begin{array}{c}{\rm Position}\\CGL_x\end{array}$&$\begin{array}{c}{\rm Momentum}\\CGL_k\end{array}$&$\begin{array}{c}{\rm Product}\\CGL_x\cdot CGL_k\end{array}$\\ 
\hline
0&1.1542&1.2350&1.4255&1.1933&1.7010&2.0299\\
1&1.1610&1.1650&1.3527&1.1599&1.3488&1.5645\\
2&1.1712&1.1346&1.3289&1.1673&1.2479&1.4567\\
3&1.1808&1.1167&1.3186&1.1767&1.2005&1.4126\\
4&1.1895&1.1045&1.3138&1.1856&1.1719&1.3894\\
5&1.1974&1.0956&1.3118&1.1938&1.1524&1.3757\\
[1ex]
\hline
\end{tabular}
\label{Table1}
\end{table*}

Neither position nor momentum Dirichlet and Neumann complexity $CGL$ from Eq.~\eqref{CGLdefinition1} depends on the applied voltage. It is seen directly from the corresponding expressions for the entropy, Eq.~\eqref{DirichletEntropy1}, and Onicescu energy, Eq.~\eqref{OnicescuDN1}. Their values, together with their products, which, of course, are field-independent too, are provided in Table~\ref{Table1} for several low lying states. It is seen that while the Dirichlet position complexity is a monotonically increasing function of the quantum number, for its Neumann counterpart this rule is broken as $CGL_{k_0}^N>CGL_{k_n}^N$ with $n=1,2,3$. Both momentum measures decrease with $n$ in such a way that the product $CGL_x\cdot CGL_k$ obeys the same law too:
\begin{subequations}\label{CGLinequalities1}
\begin{align}\label{CGLinequalities1_1}
CGL_{k_n}^{N,D}&>CGL_{k_{n+1}}^{N,D}\\
\label{CGLinequalities1_2}
CGL_{x_n}^{N,D}\cdot CGL_{k_n}^{N,D}&>CGL_{x_{n+1}}^{N,D}\cdot CGL_{k_{n+1}}^{N,D}.
\intertext{In addition, the Dirichlet product is smaller than the Neumann one:}
\label{CGLinequalities1_3}
CGL_{x_n}^N\cdot CGL_{k_n}^N&>CGL_{x_n}^D\cdot CGL_{k_n}^D.
\end{align}
\end{subequations}
\begin{figure}
\centering
\includegraphics*[width=\columnwidth]{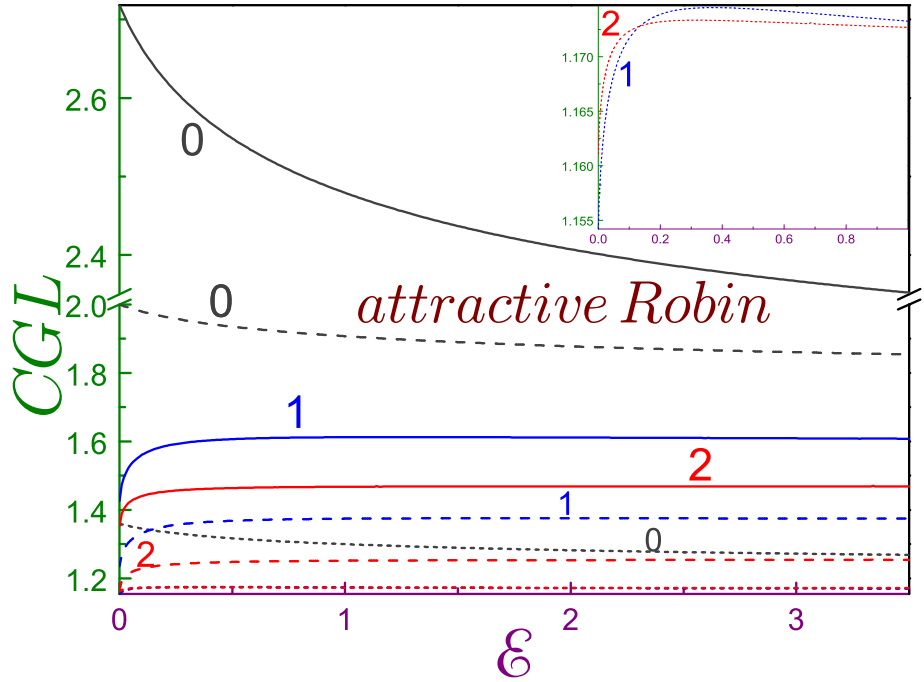}
\caption{\label{CGLfig1}
Statistical measure of complexity $CGL$ for the attractive Robin wall where dotted lines are for its position components, dashed curves -- for the momentum parts, and solid curves denote their product. Note vertical line break from $2$ to $2.35$. Numbers near the curves denote quantum number $n$. Position measures for the states $n=1$ and $n=2$ are very close to each other and  since they are not resolved in the main figure, the inset depicts them in the larger scale.}
\end{figure}
Fig.~\ref{CGLfig1} shows a complexity of the Robin attractive wall for several low-lying levels. The ground-state position and momentum measures monotonically decrease from their zero-field values, Eqs.~\eqref{CGL1}, to approach asymptotically at the high voltages those corresponding to the lowest Neumann level from Table~\ref{Table1}. For the field-induced bound state with the number $n\ge1$ a transformation from the zero-voltage value that corresponds to the  Dirichlet level with the number $n-1$ to the Neumann configuration for the quantum number $n$ is characterized by the nonmonotonicity not only of the the product $CGL_{x_n}^{R-}\cdot CGL_{k_n}^{R-}$, as it was the case for the Fisher information, but also by each of the multipliers too. As the inset demonstrates, the maxima of the complexities, which are achieved at the different weak fields for the position and momentum components and their product, are quite broad and gentle, and they flatten for the larger $n$. 

\section{Conclusions}\label{Sec_Conclusions}
Bound states of the quantum structures with non-Dirichlet BC do not necessarily satisfy the Heisenberg uncertainty relation, Eq.~\eqref{Heisenberg1}. This fact that has been known for a while \cite{AlHashimi1,Bialynicki2} has been confirmed here for the Robin surface with the negative extrapolation length $\Lambda$. Contrary to the position and momentum standard deviations from Eqs.~\eqref{DeltaXK1}, their quantum information entropy counterparts $S_x$ and $S_k$, Eqs.~\eqref{Entropy1}, always satisfy inequality from Eq.~\eqref{EntropicInequality1}. This dictates the necessity of their study for different systems, especially when the inequality \eqref{Heisenberg1} is violated. Such quantum analysis has been performed above for the interface satisfying the BC from Eq.~\eqref{Robin1} in the presence of the electric field $\mathscr{E}$ that pushes the charged particle to the wall. In addition to the quantum entropies, the properties of the position and momentum Fisher informations, Eq.~\eqref{Fisher1D_1}, and Onicescu energies, Eq.~\eqref{Onicescu1}, together with the product $e^SO$ have been calculated too. On this simplest example, a theoretical investigation of the interplay between the BC and the applied voltage discovered that the weak electric intensity transforms the Robin edge requirement basically into the Dirichlet one while for the strong fields it turns into the Neumann BC.  For these two limiting interface demands the position (momentum) quantum information entropy changes as a negative (positive) natural logarithm of the electric intensity what makes their sum a field-independent quantity that for the fixed quantum number $n$ is greater for the former BC. Similarly, the products of the position and momentum components of the Fisher information and Onicescu energy do not depend on the applied voltage either since each of them varies with the field as its direct (position) or inverse (momentum) power of one and two thirds, respectively. Contrary to the Shannon entropy and Fisher information, the Onicescu energies are decreasing functions of the quantum number $n$. For the finite nonzero $\Lambda$, these quantum-theoretical measures change their behavior from the Dirichlet-like at the low voltages to the Neumann BC at the strong electric intensities. As a result, the sum of the components of the quantum entropy and the products of the Fisher informations and Onicescu energies do depend on the field; for example, with the varying $\mathscr{E}$ the sum of the two entropies ${S_x}_0+{S_k}_0$ of the ground state attractive Robin wall crosses its counterpart of the lowest field-induced level. It was also noted that the products of the position and momentum Fisher informations of all but the zero-voltage bound state are nonmonotonic functions of the field having $n$ dependent maximum at the small $\mathscr{E}$. The energy spectrum can be efficiently controlled by the applied voltage; for example, for the attractive Robin surface at the small fields it presents  a quasi continuous part where the tiny energy difference between the levels decreases with the increase of the quantum number $n$ and, additionally, downward-split single bound state separated from the nearest upper lying counterparts by the unity gap. Spectacular thermodynamic consequences of such energy arrangement are discussed in the following paper \cite{Olendski5}.

\end{document}